# EXPLORING THE SOLAR POLES
## *THE LAST GREAT FRONTIER OF THE SUN*

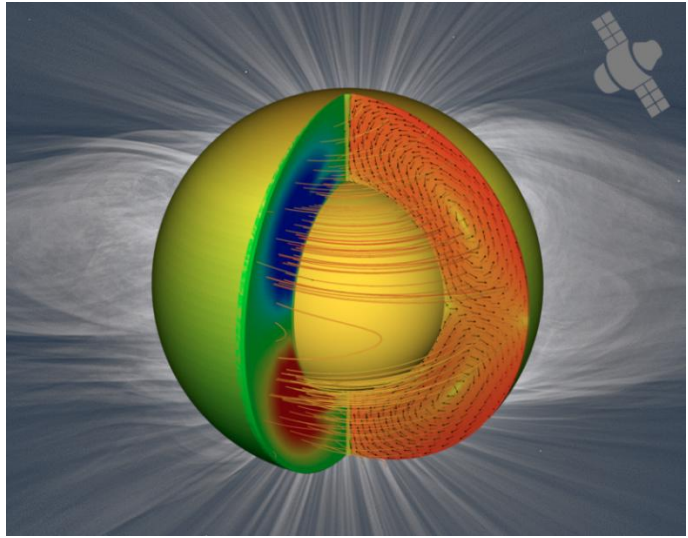

**Primary Author:** Dibyendu Nandy[1*]
**Co-authors:** Dipankar Banerjee[2], Prantika Bhowmik[3], Allan Sacha Brun[4], Robert H. Cameron[5], S. E. Gibson[6], Shravan Hanasoge[7], Louise Harra[8], Donald M. Hassler[9], Rekha Jain[10], Jie Jiang[11], Laurène Jouve[12], Duncan H. Mackay[13], Sushant S. Mahajan[14], Cristina H. Mandrini[15], Mathew Owens[16], Shaonwita Pal[1], Rui F Pinto[4,12], Chitradeep Saha[1], Xudong Sun[17], Durgesh Tripathi[18], Ilya G. Usoskin[19]

[1*]*Center of Excellence in Space Sciences India, Indian Institute of Science Education and Research Kolkata, India*, [2]Aryabhatta Research Institute of Observational Sciences, India, [3]Department of Mathematical Sciences, Durham University, UK, [4]Dept. d'Astrophysique/AIM, Université Paris et Paris-Saclay, France, [5]Max Planck Institute for Solar System Research, Germany, [6]NCAR/HAO, USA, [7]Tata Institute of Fundamental Research, India, [8]Physikalisch-Meteorologisches Observatorium Davos/World Radiation Center, Switzerland, [9]Southwest Research Institute, Boulder, USA, [10]School of Mathematics and Statistics, University of Sheffield, UK, [11]School of Space and Environment, Beihang University, Beijing, China, [12]IRAP, Université de Toulouse, France, [13]University of St Andrews, UK, [14]W. W. Hansen Experimental Physics Laboratory, Stanford University, Stanford, CA, USA, [15]Instituto de Astronomía y Física del Espacio, Buenos Aires, Argentina, [16]Dept. of Meteorology, University of Reading, [17]Institute for Astronomy, University of Hawai'i at Mānoa, [18]Inter-University Centre for Astronomy and Astrophysics, India, [19]University of Oulu, Finland

**SYNOPSIS**

Despite investments in multiple space and ground-based solar observatories by the global community, the Sun's polar regions remain unchartered territory – the last great frontier for solar observations. Breaching this frontier is fundamental to understanding the solar cycle – the ultimate driver of short-to-long term solar activity that encompasses space weather and space climate. Magnetohydrodynamic dynamo models and empirically observed relationships have established that the polar field is the primary determinant of future solar cycle amplitude. Models of solar surface evolution of tilted active regions indicate that the mid-to-high latitude surges of magnetic flux govern the dynamics leading to the reversal and build-up of polar field. Our theoretical understanding and numerical models of this high latitude magnetic field dynamics and plasma flows – that are a critical component of the sunspot cycle – lack precise observational constraints, currently limited by large projection effects due to our location in the plane of the ecliptic. This limitation compromises our ability to observe the enigmatic kilo-Gauss polar flux patches and to quantitatively constrain the polar field distribution at high latitudes. By extension, the lack of these observations handicap our understanding of how high latitude magnetic fields power polar jets, plumes, and the fast solar wind that extend to the boundaries of the heliosphere and modulate solar open flux and cosmic ray flux within the solar system. Accurate observation of the Sun's polar regions, therefore, is the single most outstanding challenge that confronts Heliophysics. A solar polar exploration mission, in isolation, or in conjunction with multi-vantage point observations across the inner heliosphere, stands to revolutionize the field of Heliophysics like no other mission concept has – with relevance that transcends spatial regimes from the solar interior to the heliosphere. This white paper argues the scientific case for novel out-of-ecliptic observations of the Sun's polar regions, in conjunction with existing, or future multi-vantage point heliospheric observatories.

**THE BIG PICTURE**

The genesis of solar-stellar magnetic activity can be traced back to a magnetohydrodynamic (MHD) dynamo mechanism operating in their interior, where plasma flows and magnetic fields interact through complex processes to generate large-scale magnetic fields (Charbonneau 2020; Brun et al. 2015). Magnetic active regions generated by the solar dynamo emerge through the Sun's surface (Jouve, Brun and Aulanier 2018) to its outer atmosphere, where subsequent dynamical interactions often lead to energetic, transient events such as solar flares and coronal mass ejections (CMEs) – which create severe space weather (Schrijver 2015). The emergence and evolution of solar surface magnetic fields, mediated via near-surface flux transport processes result in the redistribution of the surface fields which govern the structuring and dynamics of the large-scale solar corona. Solar wind is born here and propagates throughout the heliosphere. The combined action of surface field mediated open flux evolution and turbulent solar wind transport modulates the open flux, and consequently the cosmic ray flux in the heliosphere; these define the ambient space environment of solar system planets such as the Earth, upon which solar magnetic storms act as transient perturbations governing space weather. Understanding and predicting space weather is critical for protection of our space-based assets (Schrijver 2015, Daglis et al. 2021). Uncovering the influence of long-term solar magnetic variability (Usoskin 2017) and its forcing on planetary magnetosphere-atmosphere systems have profound implications for space climate and planetary habitability in the solar system and other stellar-(exo)planetary systems (Nandy et al. 2021).

While many advances have been made in these directions in the last couple of decades, the generated knowledge have also exposed critical shortcomings; the foremost being our inability to observationally constrain and fully comprehend magnetic field-associated dynamic in the Sun's polar regions and its flux content – whose influence transcends processes spanning the Sun's interior, its atmosphere and the heliosphere (Petrie 2015). ***We note that humanity has never accomplished spatially resolved precise observations of the polar magnetic fields of any star on a routine basis.*** The Solar Orbiter mission is half-step in that direction, but that is not enough. In what follows, we inspire an expedition to the Sun's polar regions focusing on some of the transformative knowledge this can generate; we do so, with the awareness that there might be new discoveries and surprises that we cannot even anticipate now.

**SUN'S POLAR FIELDS AND SUNSPOT CYCLE PREDICTIONS**

While the solar cycle governs the occurrence probability of severe space weather events and the decadal-scale forcing of planetary environments, predicting future cycles had remained an outstanding challenge (Petrovay 2018). Analyses of recent progress demonstrate that the polar field during cycle minima is the best indicator of the future solar cycle amplitude and that physics-based dynamo models based on the Babcock-Leighton mechanism have converged to indicate a weak-moderate cycle 25 (Nandy 2021). The Babcock-Leighton mechanism involves the processes of emergence of active regions (tilted bipolar sunspot pairs) and the subsequent decay, dispersal and large-scale separation of opposite polarity flux via near-surface plasma transport processes such as meridional circulation, turbulent diffusion and pumping (Jiang et al. 2014). Analytic theory indicates that surface fields are the primary driver of the internal dynamo (Cameron and Schüssler 2015). In particular, cross-equatorial cancellation and mid-to-high latitude surges from active latitudes (apparent in Figure 1d - e) are the primary determinants of solar polar field amplitude. In turn the solar polar field distribution (which determines the dipole moment) at solar minimum is the main contributor to the amplitude of the future sycle. These dynamics are currently poorly constrained by observations.

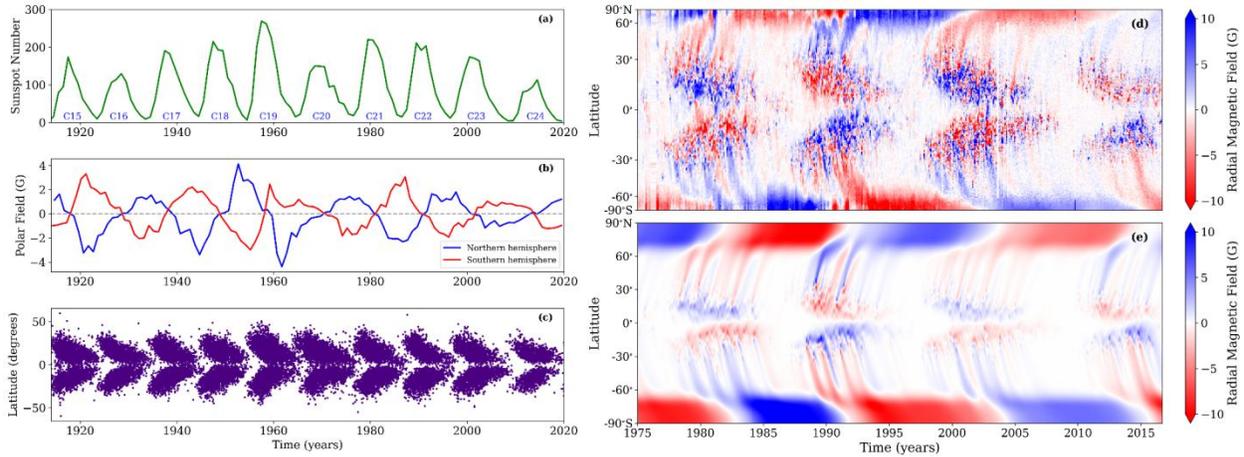

*Figure 1:* Left: The top panel shows the observed sunspot number time series over the last century, the middle panel shows the hemispheric polar field variation reconstructed from polar faculae data and Wilcox Solar Observatory data (blue: north, red: south) and the bottom panel depicts the solar butterfly diagram. Right: The top panel depicts the observed evolution of the radial field on the solar surface (gleaned from Kitt Peak Vacuum Telescope (KPVT), Michelson Doppler Imager (MDI) and Helioseismic and Magnetic Imager (HMI) observations) while the bottom panel depicts the radial field evolution as simulated from a data-driven solar surface flux transport model (Bhowmik and Nandy 2018). The figure on the left drives home the point that the polar field amplitude preceding a sunspot cycle determines the strength of the latter. The figure on the right illuminates the role of propagating "tongues" (surges) of magnetic flux from mid-to-high latitudes in the reversal and build-up of the Sun's polar field.

We do not have direct observations of the polar field and associated dipole moment variation and have to rely on proxies such as the polar faculae for validating predictive surface flux transport and dynamo models. Even more debilitating is the fact that current polar field observations suffer from large-projection effects. Figure 2 succinctly demonstrates what we can observe from a location with a direct view of the Sun's poles compared to the compromised view from plane-of-ecliptic.

***To summarize, a polar mission capable of imaging magnetic fields from at least 60 degrees above or below the ecliptic plane can return transformative information on high latitude magnetic field dynamics, constrain the polar field***

*distribution like never before, leading to the validation of magnetic field evolution models and accurate data-driven predictions of the sunspot cycle.*

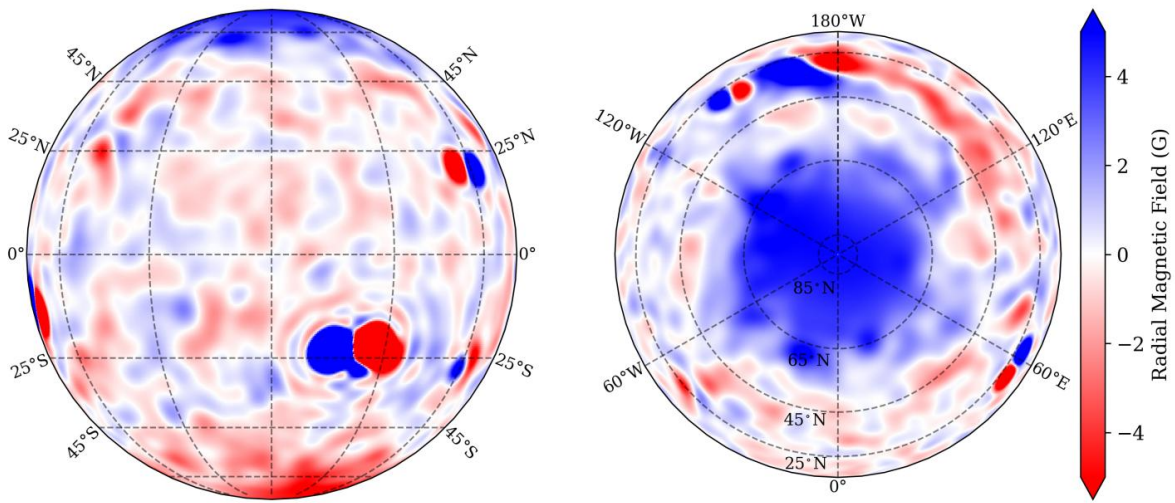

*Figure 2:* Left: A snapshot of the surface distribution of magnetic fields from a solar surface magnetic field evolution model depicting a perspective close to the plane-of-the-ecliptic (solar obliquity ignored). Right: The equivalent perspective from the top of the Sun's North pole reveals details of the polar cap missing from the low-latitude perspective. A mission that can navigate to at least 60 degrees above (or below) the ecliptic plane would reveal these details, constrain the high latitude dynamics and return accurate measurements of plasma flows and magnetic field distribution at the poles that govern solar activity and power fast solar winds.

**CONSTRAINTS ON HIGH LATITUDE SOLAR PLASMA FLOWS**

In the high beta plasma domain in near-surface layers and within the convection zone, plasma flows drive the magnetic field dynamics. These flows remain poorly constrained in the polar regions which high-latitude measurements of the surface velocity field have the potential to overcome (Löptien et al. 2015). In particular, Dopplergram data would allow us to apply helioseismic methods (Christensen-Dalsgaard, 2002; Gizon and Birch, 2005) to infer solar internal structure and dynamics. Meridional circulation – which plays a critical role in setting the timescale and amplitude of large-scale solar dynamo action and magnetism (Rempel 2006; Hanasoge, 2022) is rather poorly constrained at the poles. Significant improvements in flow measurements in the vicinity of the polar cap, including the polar extent of the meridional circulation are possible with the magnetic feature tracking technique (Mahajan et al. 2021) applied to observations that do not suffer from projection effects. Such high-latitude observations, when used concurrently with data from existing observatories such as the Helioseismic and Magnetic Imager (HMI), can generate definitive constraints on the polar meridional flow. Such measurements open up the possibility of (stereoscopic) helioseismic imaging of the plasma flows at deeper layers with greater fidelity; the deep component of the meridional flow is thought to play a critical role in the equatorward propagation and latitude of emergence of sunspots (Nandy and Choudhuri 2002), but its structure and variability remain ill-constrained. Solar internal rotation measurements – one of the greatest successes of helioseismology, is quite noisy in the vicinity of the poles.

***High latitude flows, polar vortices and associated weak rotation, and hints of significant subsurface variability of rotation are among outstanding issues that polar observations can resolve; these new constraints on rotation and meridional circulation in the polar regions can generate powerful constraints on the interplay of flow and fields that sustain solar dynamo action.***

**UNRAVELLING THE ENIGMA OF KILO-GAUSS POLAR FLUX PATCHES**

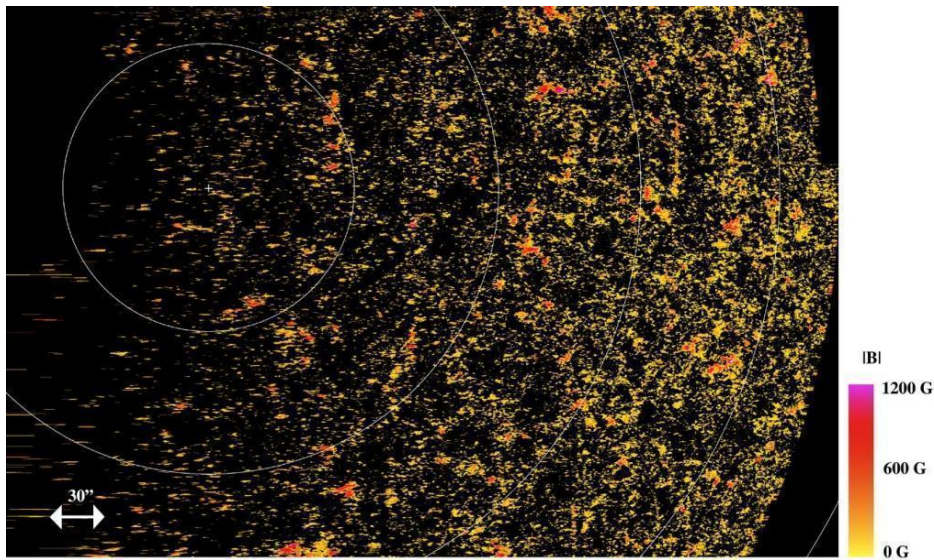

*Figure 3: Hinode SOT observations of unipolar, kilo-Gauss flux patches near the solar north pole acquired on 2007 September 25 (from Ito et al. 2010). The field of view extends over 20 degrees of the north polar cap.*

Observations by the Hinode Solar Optical Telescope (SOT) have uncovered the existence of intense small-scale kilo-Gauss flux patches in the high latitude regions of the Sun, often co-existing within, and having the same polarity as the large-scale polar cap (Tsuneta et al. 2008, Ito et al. 2010; red regions in Figure 3). These unipolar flux patches in the polar region appear to be different in nature compared to small-scale mixed polarity flux patches in the quiet Sun, and are believed to sustain open magnetic fields lines in the solar corona which could power the fast solar winds. These unipolar patches contribute significantly to the overall flux content in the polar cap and therefore may participate in the dynamo mechanism which converts the poloidal field component to the toroidal field component of the following cycle. These intense flux concentrations are also relevant in the context of the solar open flux problem (Linker et al. 2017; Wang et al. 2022), namely, a mismatch between the expected and observed open flux. The origin of these polar kilo-Gauss flux patches, the evolution of their flux content through the solar cycle, and their relation to the high latitude propagation of magnetic fields leading to polar field build up and reversal remain poorly understood.

***High latitude vantage point observations of magnetic fields and flows at different phases of the solar cycle are crucial to deciphering the mystery surrounding the origin of these polar kilo-Gauss flux patches and in conjunction with in-situ solar wind characterization – establish whether they are fundamental to the origin of fast solar winds from the poles.***

**DYNAMICS OF THE POLAR ATMOSPHERE: JETS, PLUMES AND WINDS**

The Sun's poles are often characterized by the presence of coronal holes, open solar magnetic field lines, intense kilo-Gauss flux tubes, polar plumes and jets (Hanaoka et al. 2018), and sometimes pseudo-streamers. Open coronal field lines, ubiquitous in the polar regions are excellent pathways for Alfvén wave (turbulent dissipation) mediated solar wind acceleration (Hassler et al. 1999; Krishna Prasad, Banerjee and Gupta 2011; Morton, Tomczyk and Pinto 2015). The strength of the solar polar field during different phases of the cycle and at minima influences the boundaries of open and closed field lines and the tilts of large-scale streamers (Dash et al. 2019). Recent findings hint that Interchange reconnection in coronal holes may power the solar wind; however, proving this would require polar magnetic field measurements (Tripathi, Nived and Solanki 2021; Upendran and Tripathi 2022). All of these polar phenomena play a role in the acceleration and global structuring of fast solar winds, which leave their imprint in the heliosphere and near-Earth space weather (Cranmer, Gibson and Riley 2017).

***Adequate constraints on the landscape of the polar magnetic fields and transient phenomena through the solar cycle – achieved via an out-of-ecliptic space mission concept designed for multiple passes over the solar poles – would transform our ability to assess and predict the large-scale structuring of the global solar winds and their heliospheric and space weather consequences, and bring to the fore magnetic connectivity bridging the solar interior to the farthest reaches of the heliosphere.***

**SUMMARY RECOMMENDATIONS**

***NASA Programs:*** Addressing the novel science presented here necessitates space mission whose scope may either be Moderate-scale to Large-scale depending on whether a single-vantage point, or a multi-vantage point mission concept is planned to address the science. We believe that a proof-of-concept, single-vantage point mission with a quick turn-around time for imaging the magnetic fields of the polar regions and solar atmosphere from at least 60 degrees above the plane-of-the ecliptic during the high latitude dynamics leading up to the next solar minimum (i.e., 2026-2032) time frame, with a nominal mission lifetime of 5-10 years would be a prudent approach (e.g., whitepaper by Hassler et al. 2022). Concurrently, and with the experience of this mission assimilated, a multi-vantage point mission concept may be explored to constrain solar activity and in-situ space environment across the inner heliosphere.

***NSF Programs:*** Theory, numerical simulations, and data analysis of existing observations from ground and space that directly contribute to a holistic understanding of physical processes encompassing the solar interior, atmosphere and the space environment of solar system planets, with a view to establishing causality are necessary elements for community preparedness and the eventual success of future space missions. We believe that mid-scale Research Infrastructure and Innovation programs supported by NSF are therefore crucial to complement any NASA space mission. Specifically, we propose supporting large teams or Centers that leverage and assimilate US and global expertise to develop the theoretical, modeling and data analysis tools that will stand to complement and enable transformative scientific returns from the out-of-ecliptic, multi-vantage point space missions envisaged in this white paper.

***Globally Coordinated Multi-Vantage Point Observations of the Heliosphere:*** We note that existing missions such as Parker Solar Probe, Solar Orbiter, Solar Dynamics Observatory, ACE, WIND, DSCOVR and upcoming missions such as CNSA Advanced Space-based Solar Observatory, ISRO's Aditya-L1 and ESA's Vigil offer the possibility of globally coordinated observations in conjunction with a novel solar polar mission. We recommend that this be taken full advantage of by cementing mechanisms for data exchange across mission teams and leveraging modeling and data analysis expertise across borders. This synergy may catalyze more ambitious, multiagency international programs for system wide exploration of Heliophysics through the deployment of multiple spacecrafts to cover the whole 4π steradian of the heliosphere (e.g., whitepaper by Raouafi et al. 2022) that may be prohibitively expensive for any one nation alone; this should be the ultimate goal for humanity, to together understand the space we share and use this understanding to guide explorations of other worlds, beyond our own.